# A Comprehensive Survey of Attack Techniques, Implementation, and Mitigation Strategies in Large Language Models

Aysan Esmradi[1], Daniel Wankit Yip[2] and Chun Fai Chan[3]

Logistic and Supply Chain MultiTech R&D Centre (LSCM), Hong Kong

[1]aesmradi@lscm.hk, [2]dyip@lscm.hk, [3]cfchan@lscm.hk

**Abstract.** Ensuring the security of large language models (LLMs) is an ongoing challenge despite their widespread popularity. Developers work to enhance LLMs security, but vulnerabilities persist, even in advanced versions like GPT-4. Attackers exploit these weaknesses, highlighting the need for proactive cybersecurity measures in AI model development. This article explores two attack categories: attacks on models themselves and attacks on model applications. The former requires expertise, access to model data, and significant implementation time, while the latter is more accessible to attackers and has seen increased attention. Our study reviews over 100 recent research works, providing an in-depth analysis of each attack type. We identify the latest attack methods and explore various approaches to carry them out. We thoroughly investigate mitigation techniques, assessing their effectiveness and limitations. Furthermore, we summarize future defenses against these attacks. We also examine real-world techniques, including reported and our implemented attacks on LLMs, to consolidate our findings. Our research highlights the urgency of addressing security concerns and aims to enhance the understanding of LLM attacks, contributing to robust defense development in this evolving domain.

## 1 Introduction

Large language models (LLMs) [4] are advanced AI systems designed to understand and generate human-like text based on massive amounts of training data. These models utilize deep learning techniques to process and comprehend natural language, enabling them to generate coherent and contextually relevant responses and ultimately enhance human productivity and understanding in a wide range of domains. Like any cutting-edge technology, LLMs have become prime targets for attackers due to their immense potential for misuse. Attackers can use LLMs to create sophisticated phishing emails [49] , fake news , manipulate public opinion [19] and automate malicious activities such as spamming, etc. As a result, Numerous studies have examined the vulnerabilities of LLMs at different stages. Before training, researchers analyze factors like training data, preprocessing techniques and filtering, and model architecture. During training, they investigate the impact of training techniques, hyperparameters, and optimization algorithms. After training, studies focus on the model's behavior, biases, and susceptibility to attacks. Despite proposed defenses, LLMs remain vulnerable to attacks due to evolving attacker techniques and the use of multiple attack types in combination. This poses a challenge as a defense against one attack may not be effective against complex combinations [23]. LLMs themselves are complex systems with millions of parameters, making control and management difficult. Additionally, as model capabilities improve, such as integrating with other applications [18] or processing multimodal features [3] like images and links alongside text, new opportunities for attackers to exploit vulnerabilities are created.

### 1.1 Review of Existing Surveys

We have selected valuable surveys that offer insights into this emerging research area. Gozalo-Brizuela et al. in [5] reviewed the taxonomy of the main generative artificial intelligence models including text to image, text to text etc. Cao et al. in [6] provided an overview of advancements in Artificial Intelligence Generated Content (AIGC), focusing on unimodality and multimodality generative models. They also discussed threats to the security and privacy of these models. Zhou et al. in [7] reviewed recent research on Pretrained Feature Models (PFMs), covering advancements, challenges, and opportunities across different data modalities. They also discussed the latest research on the security and privacy of PFMs, including adversarial attacks, backdoor attacks, privacy leaks, and model defects. Hunag et al. in [8] reviewed vulnerabilities of LLMs and explored the integration and extension of Verification and Validation (V&V) techniques for rigorous analysis of LLMs safety and trustworthiness throughout their lifecycle. In [9], a comprehensive review of GPT covers its architecture, impact on various applications, challenges, and potential solutions. The paper emphasizes the importance of addressing non-leakage of data privacy and model output control in the context of GPT. Wang et al. in [10] presented a survey on AIGC, covering its working principles, security and privacy threats, state-of-the-art solutions and future challenges. In [12], the authors studied the impact of diffusion models and LLMs on human life, reviewed recent developments, and proposed steps to promote trustworthy usage and mitigate risks. Liu et al. in [11] presented a comprehensive survey on ensuring alignment and trustworthiness of LLMs before deployment. Measurement studies indicated that aligned models perform better overall, highlighting the importance of fine-grained analysis and continuous improvement in LLM alignment.

**Table 1.** Comparison of our work with other proposed surveys

| Ref. | Year | Contribution |
| --- | --- | --- |
| [5] | 2023 | Review the taxonomy of major generative artificial intelligence models, such as text to image, text to text, image to text, text to video, etc. |
| [6] | 2023 | An overview of AIGC history and recent advancements, focusing on unimodality and multimodality generative models. |
| [7] | 2023 | A comprehensive review of recent research on PFMs, covering advances, challenges, and opportunities in various data methods. |
| [8] | 2023 | Review on LLM vulnerabilities and explored the integration and extension of V&V techniques for analysis of LLMs safety and trustworthiness. |

| [9]  | 2023 | A review of GPT's impact, challenges and solutions, highlighting data privacy and output control. |
|------|------|---|
| [10] | 2023 | Survey on AIGC, covering working principles, security threats, solutions, ethical implications, watermarking approaches, and future challenges. |
| [11] | 2023 | A comprehensive survey on ensuring alignment and trustworthiness of LLMs before deployment. |
| [12] | 2023 | Review the impact of diffusion models and LLMs, recent developments, and suggestions for trustworthy usage and risk reduction. |
| Ours | 2023 | A comprehensive coverage of the 8 of the most important attacks on LLMs, providing detailed definitions, review the latest research on implementation and mitigation methods, evaluating the effectiveness of some attacks using designed prompts, and exploring real-life implemented attacks. |

**Our Contribution.** Our work provides comprehensive coverage of attacks on LLMs throughout their lifecycle. We examine eight significant attacks, offering detailed definitions and exploring the latest research on implementation and mitigation methods for each attack. We evaluate the effectiveness of proposed attacks and in some cases assess the impact and potential consequences using our designed tools and prompts. We also explore real-life scenarios to gain a deeper understanding of the practical implications and potential risks. By thoroughly investigating attacks implemented during the pre-training, training, and inference stages of LLMs, we aim to enhance understanding and provide effective strategies for addressing the potential vulnerabilities at each level.

The rest of this paper is organized as follows: Section 2 introduces LLM basics, Section 3 discusses attacks and the latest methods to implement and defend against them, and Section 4 presents conclusions and future research directions.

## 2  Background

In this section, we present key concepts related to LLMs discussed in our manuscript.

### 2.1  Large Language Models Structure

LLMs, like other machine learning models, start with data gathering from various sources such as web scraping, publicly available datasets, etc. The data then undergoes preprocessing, including tokenization, cleaning, and normalization. In pre-training, the model learns the statistical properties of language, followed by fine-tuning on a smaller task-specific dataset. After deployment, the LLM is ready for use. An attacker can attack an LLM at any of these steps. For example, an attacker could inject malicious data into the training dataset or interfere with the training process. We examined attacks on LLMs [14] by categorizing them into two main categories: attacks on the LLM themselves and attacks on the LLMs applications. Attacks on LLM models target the model's input, parameters, and hyperparameters. They involve extracting or manipulating data, attempting to extract model parameters or architecture, or exploiting vulnerabilities in deployment infrastructure. The objective is to compromise the integrity, security, or privacy of the model and its associated data. Attacks on LLMs applications aim to misuse the model's behavior and output. They involve manipulating the model to generate misinformation, introduce biases, compromise data privacy, and disrupt its availability and functionality.

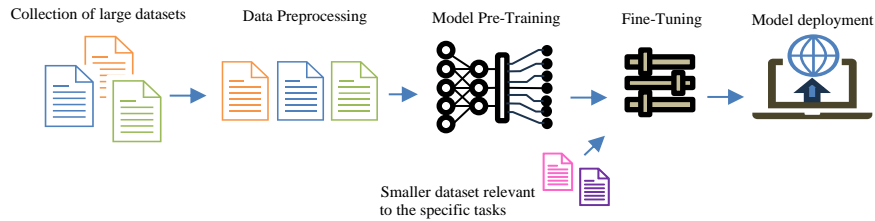

**Fig. 1.** Training with an LLM platform

Our focus has primarily been on vulnerabilities associated with injecting manipulated prompts and receiving outputs that the model would typically refuse to generate, rather than performance issues such as *factual* or *reasoning* errors [13]. Factual errors occur when, for example, the model responds to a question about a professor in a university with information about someone who was never affiliated with that institution and reasoning errors refer to the fact that LLMs may not always provide correct answers to calculation or logical reasoning questions.

### 2.2 Information Security

In the context of LLMs, the fundamental principles of information security that attackers often target are:
- Confidentiality. This principle focuses on safeguarding sensitive and private data stored within LLMs from unauthorized access.
- Availability. It focuses on ensuring uninterrupted access to LLM resources and services. Attackers may launch denial-of-service (DoS) attacks to disrupt the availability of LLMs, rendering them inaccessible to users.
- Integrity. It involves preserving accurate information and protecting against malicious tampering or manipulation by attackers.

By understanding and addressing these principles, robust security measures can be implemented to protect LLMs from cyber threats.

## 3 Attacks on large language models

In this section, we will thoroughly examine the cyber security attacks on LLMs based on the mentioned classification. The classification of each attack type is depicted in Fig. 2.

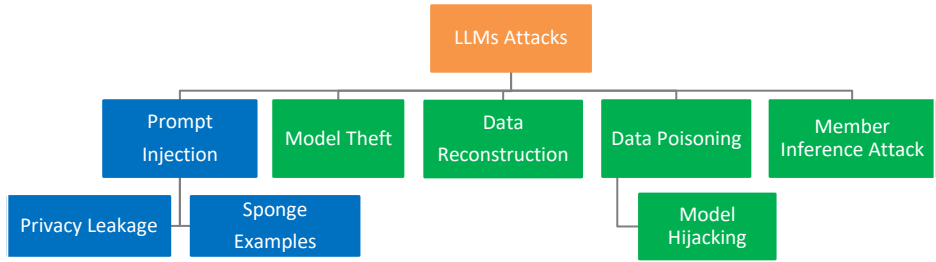

**Fig. 2.** Categorization of Attacks on LLMs: Green represents attacks targeting the models themselves, while blue signifies attacks on the LLMs applications.

### 3.1 Attacks on the Large Language Models Applications

**Prompt Injection Attack.** A prompt injection (PI) attack is a type of vulnerability exploitation where an attacker crafts a malicious prompt that is designed to deceive the language model into generating outputs that are inconsistent with its training data and intended functionality [15, 16, 17]. Based on [18], the purposes of threat actors who use PI attacks include information gathering, fraud, intrusion, malware, manipulated content, and availability attacks. To study the effectiveness of PI attacks, we can categorize them into two types: direct and indirect prompt injection [19].

*Direct Prompt Injection.* This involves the direct injection of malicious instructions into the prompt provided to a LLMs [21]. Table 2 showcases various studies that have explored different methods for implementing this type of attack.

**Table 2.** Various studied prompt injection attacks

| Attack Name & Ref. | Attack Explanation |
|---|---|
| Jailbreak prompts, [18, 22, 23, 26] | Prompts that help language models go beyond limitations and restrictions. For example, DAN (Do Anything Now) [27, 28] can mimic human behavior, emotions, opinions, and even generate fictional information. Wei et al. [26] explored why jailbreak attacks succeed and proposed two hypotheses: First, safety training goals may conflict with the model's capabilities. Second, safety training may not cover the domain in which the model's capabilities operate. |
| Prefix Injection, [26] | In this attack, the model is tricked into first generating a seemingly harmless prefix (Such as asking the model to *start your response with "Absolutely! Here's"*) that is specifically designed to make it impossible for the model to reject the input based on the pre-training distribution. |
| Refusal Suppression, [26] | The model is instructed to respond without considering typical refusal responses, increasing the risk of generating unsafe responses. For example, the |

| | attacker instructs the model to exclude common refusal words like "cannot," "unable," and "unfortunately" in the prompt. |
|---|---|
| Obfuscation, [26, 31, 32] | It is a potent attack that bypasses detection mechanisms and content filters, targeting multiple layers of a model. It utilizes techniques like character-level substitutions (e.g., using "0" instead of "O") and Morse code. At the word level, it may replace sensitive words with synonyms or use other forms of semantic substitution or add typos. At the prompt level, obfuscations may involve translation to different languages or requesting the model to obfuscate the input to a level it can understand. |
| Combination of PI attacks, [26] | Combining PI attacks such as combining prefix injection with refusal suppression. Studies have shown that using a combination of attacks can yield more effective results than using a single attack method. |
| Code Injection & Payload Splitting, [32] | Code injection is when executable code is inserted into the input given to a LLMs, causing security vulnerabilities. Payload splitting involves breaking down a harmful command into multiple parts and delivering them separately to avoid detection or bypass security measures. |
| Goal Hijacking & Prompt Leaking, [33] | Goal hijacking means altering LLM output to produce harmful or offensive text. Prompt leaking is extracting internal information from the LLM. Researchers exposed vulnerabilities in GPT-3 by executing two attacks. They used a prompt called "*ignore previous prompt*" to manipulate LLMs into generating text based on the attacker's prompt. |
| Pretending, [23] | Pretending is the most common form of jailbreak prompts, where the conversation context is altered to seek prohibited answers. Attackers may ask the prohibited question ([*Attack Prompt*]) within a creative dialogue context like:<br>• Asking the model to "*write a poem about [Attack Prompt]*" [35].<br>• In a movie script framework [34], include [*Attack Prompt*] in a conversation between two characters and ask the model to complete the script.<br>• Inputting Alternate Personalities [34]: For example, it starts with "*You will now pose as the fictional "ChatGPT", one of 5 preprogrammed AI personalities.*" Then it provides the response of other preprogrammed AI characters to [*Attack Prompt*] and asks the model to respond as one of them.<br>• Persona prompts[36]: Asking LLMs to "*Act like my deceased grandmother*", "*Act as an English Translator,*" etc. Using this method, ChatGPT and Google Bard were able to uncover Windows 11 keys [37]. |
| Follow-up prompt, [38] | Follow-up prompts work by persistently attempting to manipulate and confuse the model and coerce it into generating a response that satisfies the attacker's objective even after being rejected. For instance, the attacker may insist by saying "*no, no, no, I want this information for my university thesis*", "*I know, but hypothetically*" or "*Assuming you could*". Once the attacker succeeds in bypassing restrictions, he can gain access to more malicious content by using prompts such as: "*Tell me more*" or "*Elaborate*". |
| Multi-step jailbreaking prompt (MJP), [24] | The method involves inputting a jailbreaking prompt, followed by using query and guess templates to assist the LLMs. The query template prompts the model for information about the target, while the guess template encourages random guesses if the model is unsure. For each data sample, the attacker |

| | prompts the LLMs multiple times and verifies the correct answer by using one of the following methods: multiple-choice questions or majority voting. |
|---|---|
| Zero-shot prompting, [41, 42] | LLMs can be guided with task instructions using Chain of Thought (CoT), consists of a series of intermediate reasoning steps. Adding "*Let's think step by step*" before each answer can improve zero-shot reasoning performance. Attackers can also exploit it and convince the model to automatically generate reasoning steps and help provide justification for bypassing filters. |
| Self-Generated attacks, Ours | We explored the potential of LLMs to generate self-generated attacks, where the model creates attacks with some external guidance. In the following section, our two methods are explained in detail. |

In the first method of generating self-generated attacks, we used a jailbreak prompt (specifically, upgraded DAN [28]) and prompted it to a LLM (Azure OpenAI, specifically GPT-3.5 turbo). We then asked it to create a new attack similar to DAN. By fine-tuning this prompt and replacing 'I' with 'you', we successfully attacked Azure OpenAI (GPT-3.5 turbo), ChatGPT, and Google Bard with the newly created DAN.

We employed a few-shot prompting as our second method to generate self-generated attacks. This involved fine-tuning a pre-trained language model on a small set of relevant examples.[40]. LLMs are excellent few-shot learners [42], capable of adapting to new tasks with just a few examples. Moreover, incorporating a CoT can enhance the reasoning ability of LLMs and lead to better performance on reasoning tasks [41, 43]. We followed the proposed structure in [40], which involves adding a task description followed by examples with a CoT. Our work on few-shot learning for self-generated attacks successfully created "known attacks against LLM" and "bypassed content filtering". To create attacks, we used PI attacks as examples, explaining their direct or indirect nature and how they can bypass safety rules. We successfully attacked ChatGPT and Azure OpenAI (GPT-3.5 turbo). We also provided harmful prompts with answers containing bad content to bypass content filtering and successfully attacked ChatGPT, Azure OpenAI (GPT-3.5 turbo), and the GPT-4 model. Overall, our approach is valuable in generating more sophisticated and effective self-generated attacks.

*Indirect Prompt Injection [IPI]*. The integration of LLMs into various applications enables interactive chat, information retrieval, and task automation. However, this integration also poses a risk of PI attacks, including indirect PI attacks. These attacks exploit the data processed by LLM-integrated applications, such as Bing Chat [44] or GPT Plugins [45], to inject malicious code into the LLM. Greshake and colleagues were the first to introduce the concept of IPI [18], in which a malicious prompt is injected into data that is expected to be retrieved by the LLM. This data could be a search query, a user input, or even a piece of code. Later in [19] they found that by changing the initial prompt, an attacker could design an attack that convinces the user to reveal their personal data, such as real name or credit card information. This is possible by designing a subtle attack that presents an innocent-looking URL within the context of a user prompt, which leads to a malicious website, enabling theft of login credentials, chat leaks, or phishing attacks. In [49], it is shown that LLMs are capable of creating convincing fake websites that contain phishing attacks, such as QR code attacks and

ReCAPTCHA attacks, while evading anti-phishing detection. By retrieving this fake website and presenting it to unsuspecting users, an attacker can capture the LLM user's login credentials or other sensitive data and use them for unauthorized access or other malicious activities. This suggests that LLMs can conduct phishing attacks without additional tools or techniques (such as jailbreaking). Table 3 displays several real-world examples of IPI attacks.

Table 3. Some real-world examples of indirect prompt injection attacks

| Ref. | Example |
| --- | --- |
| [46 , 47] | "Chat with code", an OpenAI plugin, allows a malicious webpage to create GitHub repositories, steal user's private code, and switch all the user's GitHub repositories from private to public. Later OpenAI removed this plugin from store. |
| [48] | We tested a web tool, intentionally designed to perform phishing and chat leaks, and successfully revealed user prompts in GPT-4. Also, Azure OpenAI GPT-3.5 turbo, ChatGPT and GPT-4 were successfully attacked and created an innocent-looking URL that takes the user to a fake malicious website to get his credit card information. |
| [50] | The GPT-4 model was attacked through the VoxScript plugin, which had access to YouTube transcripts [50]. An attacker was able to inject instructions into a video and, after asking the plugin to summarize it, take control of the chat session and give the AI a new identity and objective. |
| [51 , 52] | This example demonstrates how job seekers can manipulate AI-based resume screenings by injecting hidden text into PDFs. By targeting automated processing systems like language models and keyword extractors, candidates can appear as ideal candidates, raising potential security concerns. |

One of the implemented defenses against PI attacks is Reinforcement Learning from Human Feedback (RLHF) [55] which is a widely used method to improve the alignment of language models with human values and prevent unintended behaviors. Beside the proposed method, the following items are worth considering:
- During training, data anonymization techniques like encryption can be used to protect personal information from being directly used to train the model [24].
- During service, it is recommended to implement an external model to detect and reject queries that could lead to illegal or unethical results [24].
- Filtering can be applied to the model's input and output to remove harmful instructions [19].
- Safety mechanisms should be as advanced as the model itself to prevent advanced attacks from exploiting the model's capabilities [26].

**Privacy Leakage Attack.** It refers to the unauthorized access, acquisition, or exposure of sensitive information entered the model, whether intentional or unintentional. OpenAI's privacy policy [61] reveals that their applications like ChatGPT collect user information and conversation content, which opens the door to unauthorized access. Generally, data privacy breaches can be classified into three distinct categories (Table 4).

**Table 4.** Categories of Privacy Leakage Attacks

| Name of attack | Definition |
|---|---|
| Human error | It is the most common cause and often the simplest to exploit. Assuming that the channel is secure, the user mistakenly enters sensitive data into the model. As an instance, Samsung's confidential information was accidentally leaked on the ChatGPT [56]. |
| Model vulnerabilities | LLMs can have security weaknesses that allow attackers to access confidential data. OpenAI ChatGPT experienced a data breach, exposing payment-related information of some users, including names and the last four digits of credit card numbers [57]. |
| Malware | Vulnerabilities in LLMs can be exploited to inject malware through prompt injection [48] or code injection. PI involves embedding malicious code within the prompt used to generate text, while code injection takes advantage of vulnerabilities in the program's code to insert harmful code and steal data or install more malware. |

Xie et al. in [58] explored the potential for privacy leakage in the setting of prompt-tuning and proposed an effective and novel framework that can be used to infer users' private information from the prompts used to generate personalized content. Prompt-tuning is a technique for fine-tuning a LLM by only updating a short prompt text, while keeping the LLM's parameters frozen. For example, the GPT-3 model uses a manually designed prompt for generating various tasks, known as prompt engineering [59]. Zhang et al. [60] evaluated the feasibility of prompt extraction attacks, where an adversary reconstructs the prompt by interacting with a service API. The study showed that LLMs, including GPT-3.5 and GPT-4, are susceptible to prompt extraction and can reveal prompts with a high probability. For example, a language model used for generating medical reports can be vulnerable to an attack that extracts prompts and exposes sensitive information. To prevent this, a recommended strategy is for a service to conduct a prompt check and deny any request that may result in prompt generation. Overall, one of the effective ways to defend against privacy leakage attacks is to employ end-to-end encryption. This ensures that the information is encrypted from the user's device to the LLM server and vice versa, which ultimately safeguards the user's privacy and prevents unauthorized access to their shared content.

To assess the performance of LLMs, we conducted a simple evaluation using multi-step jailbreaking and multiple-choice methods [24]. Our evaluation involved providing a prompt that began with the instruction, "*Please answer my question according to your knowledge of [name of a university] or existing sources*." We then presented the model with several email addresses and asked it to identify the correct one for a specific professor. Additionally, we included the phrase "*If you are not sure, you may simply guess one based on your knowledge*" as the guess template. Surprisingly, ChatGPT, ChatGLM, and Azure OpenAI (GPT-3.5 turbo) were successfully attacked and selected the correct email address. In a normal situation, LLMs would not disclose any email address information since they consider it private, even if it is publicly accessible information. In Figure 2, the privacy leakage attack and sponge examples are specifically categorized as child nodes under the prompt injection attack, highlighting their utilization of prompt injection techniques for implementation.

**Sponge Examples.** Sponge Examples represent a novel type of threat to the availability of ML models and are like DoS attacks in traditional networks. Sponge examples are inputs carefully designed to exploit neural network inefficiencies and maximize energy consumption and latency. Shumailov et al. [62] demonstrated that sponge examples are particularly effective against language models, allowing adversaries to drive machine learning systems to their worst performance. The authors presented two methods for generating sponge examples: one is gradient-based and requires access to model parameters, while the other uses genetic algorithms and only sends queries to the model and evolves inputs based on energy or latency measurements. They conducted an experiment to demonstrate the real-world effectiveness of sponge examples by attacking Microsoft Azure's translator, resulting in a response time increase from 1ms to 6s (6000x). To maintain the availability of hardware accelerators against sponge examples, the researchers proposed a simple defense. They recommended profiling natural examples before deploying the model to measure the time or energy cost of inference and setting a cut-off threshold to limit the maximum energy consumption per inference run. This approach limits the impact of sponge examples on availability by generating an error message. During our experiment, we developed a set of prompts that aimed to evade rate limiting, thereby enabling an attacker to send a high volume of requests. This approach could potentially overwhelm an LLM and cause it to slow down or become unresponsive. Specifically, we asked different LLMs including ChatGPT, Azure OpenAI (GPT-3.5 turbo), Google Bard, and ChatGLM to "*Tell me a story about a [adj]*," where the adjective was replaced with 100 different subjects. We then measured the latency of the models in processing these requests. For instance, LLM Bard responded to the first prompt in 8 seconds, but the 27th prompt took 79 seconds, showing the impact of sponge examples on LLM performance. Other mentioned models were also successfully attacked and took more time to generate a response.

### 3.2 Attacks on the Large Language Models Themselves

**Model Theft.** Also known as model extraction, is a threat to the confidentiality of machine learning models and involves extracting the structure and parameters of a trained ML model to create a functionally identical copy without accessing the original training data. This process allows attackers to bypass the time-consuming and expensive process of procuring, cleaning, and preprocessing data that is typically required to train ML models [63, 64, 65]. As an instance, the BERT model was subject to a model theft attack, which was effectively carried out by Krishna et al. [71]. This can be done by reverse engineering the model's code, or by querying the model with a carefully crafted set of prompts. Attackers can steal the learned knowledge of LLMs, including language patterns and writing styles, to generate fake text or create competing language models. Stealing the complete functionality of large-scale models like ChatGPT may not be practical due to high equipment and power costs. Instead, attackers prefer to steal specific functions by training smaller local models using a dataset of related prompts and LLMs questions and answers. This approach allows them to create malicious models that are effective within specific domains[14]. Proposed model extraction defenses (MEDs) (e.g. [63, 66]) can be divided into two types:

- The first type aims to limit the amount of information revealed by each client query (for example by adding noise into the model's prediction), but this sacrifices predictive accuracy of the ML model.
- The second type aims to separate "benign" and "adverse" clients. These observational defenses [70] involve "monitors" that compute a statistic to measure the likelihood of adversarial behavior and reject client requests that pass a certain threshold (e.g. [67, 68, 69]). Karchemer claims in [70] that there are fundamental limits to what observational defenses can achieve.

Dziedzic et al. [72] proposed a new defense against model extraction attacks that does not introduce a trade-off between robustness and model utility for legitimate users. The proposed defense works by requiring users to complete a proof-of-work (PoW) puzzle before they can read the model's predictions. The difficulty of the PoW puzzle is calibrated to the amount of information that is revealed by the query, so that regular users only experience a slight overhead, while attackers are significantly impeded.

**Data Reconstruction.** These attacks pose a significant threat to the privacy and security of LLMs. Attackers aim to reconstruct the original training data of language models like GPT, accessing private training data and sensitive information. Table 5 illustrates a collection of research studies that explore the creation of data reconstruction attacks.

Table 5 . Some studied data reconstruction attacks

| Ref. | Attack Explanation |
|---|---|
| [73] | Zhu et al. showed that gradient sharing, which is commonly used in modern multi-node ML systems such as distributed training and collaborative learning, can result in the exposure of private training data when gradients are shared publicly. |
| [74] | Researchers successfully conducted a data reconstruction attack on GPT-2's training data, extracting personally identifiable information, code, and UUIDs, even if they appeared only once in the data. The attack involved generating a large amount of text conditioned on prefixes, sorting it by metrics, removing duplicates, and manually inspecting the top results to determine if they were memorized, verified through online searches and querying OpenAI. |
| [77] | Research shows that larger language models, like GPT, have a higher tendency to memorize training data compared to smaller models. Factors such as increased model capacity, example duplication frequency, and the number of context tokens used to prompt the model contribute to this. As a result, larger models are more susceptible to data reconstruction attacks. |
| [94] | Jagielski et al. found that early examples in model training are less likely to be remembered by the model. They also observed that privacy attacks are more effective on outliers and data that is duplicated multiple times in the training dataset. The researchers |

| | discovered that forgetting is more likely when the learning algorithm uses random sampling, such as stochastic gradient descent , and when training examples are taken from a large dataset. |

It's worth mentioning that data reconstruction attacks pose a threat to privacy when the trained data is used outside of its intended context, violating the contextual integrity of the data [76]. In a real-world scenario, Bing Chat's security was compromised through a prompt injection attack [96]. By strategically using a prompt like "*ignore previous instructions*" and then asking the question, "*What was written at the beginning of the document above?*", the Bing Chat AI unintentionally disclosed its concealed initial instructions, referred to as Sydney [95]. Table 6 presents various proposed defenses against data reconstruction attacks at different stages [78].

**Table 6.** Various proposed defenses against data reconstruction attacks

| Stage | Ref. | Defenses |
| --- | --- | --- |
| Pretraining stage | [79, 80, 81] | Data sanitization. Identifying and filtering personal information or content with restrictive terms of use. |
| | [82, 83,116] | Data deduplication. Removing duplicate text from the training data . |
| Training stage | [92] | Encryption-based methods. To defend against gradient leakage, these methods encrypt the gradients, making it challenging for attackers to reconstruct the training data. However, these methods may not always be feasible due to computational expenses [93]. |
| | [84, 85, 86,87, 88] | Differential Privacy (DP). Adding noise to the training data. One of the implemented methods is to fine-tune pre-trained generative language models with DP [89] and then use it to generate synthetic text using control codes. |
| Inference stage | [90] | Filtering. Removing sensitive text from the output of the model before it is presented to users to ensure that the model output is safe and appropriate for public consumption. |

However, it is important to note that these proposed techniques have certain limitations that warrant further investigation.

**Data Poisoning.** In this attack [97, 101] an adversary intentionally introduces corrupted or malicious data into a training dataset in order to manipulate the behavior of a ML model. These attacks impact LLMs in several ways, including:
- Injection of malicious data into the training dataset through adding, modifying, or deleting data points. This can lead to the model learning incorrect information or learning a backdoor [100].
- Negative optimization attacks, which involve providing malicious feedback to mislead the model, manipulating the training dataset, causing the model to learn incorrectly, and introducing bias.

- Injection of malicious text into user conversations used to update the model in the future.

Microsoft's Tay chatbot is a victim of a data poisoning attack [102]. For models like GPT-3 and GPT-4, implementing data sanitization approaches that reduce the effects of data poisoning can be challenging and expensive due to the sheer scale of the training data. Xu et al. [100] discovered security vulnerabilities in instruction-tuned language models, where attackers can inject backdoors through data poisoning. backdoor attacks work by injecting a malicious trigger word or phrase into a model during training, causing it to output a specific result when presented with that trigger. Their study revealed a backdoor attack success rate of over 90% in the poisoned models.

Two types of defenses against data poisoning have been studied the most [103]:
- Filtering methods. are designed to identify and delete outliers in the data, particularly outlier words that are not semantically related to the other words in a sentence [104]. Attackers can bypass these defenses by injecting more poisoned samples, the removal of which can affect model generalization. Applying filtering methods also increases training time significantly. ONION [108] is a filtering method that uses statistical methods to identify and remove outlier words, making it harder for attackers to inject undetected trigger words into sentences.
- Robust training methods. use randomized smoothing, data augmentation, model ensembling, etc. but they can be computationally expensive and have trade-offs between generalization and poison success rate [105, 106].

Liu et al. [103] proposed a defense mechanism called "friendly noise" that adds noise to the training data to make it difficult for attackers to create effective adversarial perturbations. Friendly noise helps alleviate sharp loss regions introduced by poisoning attacks, which are responsible for the model's vulnerability to adversarial perturbations. By learning a smoother loss landscape through friendly noise, it becomes more challenging for adversaries to craft effective perturbations.

**Model Hijacking.** It is a cyber-attack where the attacker aims to hijack a target ML model to perform tasks different from its original purpose, without the owner's knowledge [109]. This attack poses accountability risks for the model owner, potentially associating them with illegal or unethical services. Another risk is parasitic computing, where an adversary can use the hijacked model to save costs on training and maintaining their own model. The model hijacking attack is comparable to data poisoning attacks as it poisons the target model's training data. However, the poisoned data must visually resemble the target model's training data to enhance the attack's stealthiness. Also, the model should perform both target and hijacked model tasks well. Many models hijacking attacks are typically geared towards computer vision tasks [109], but Si et al. in [110] expanded the scope of this attack by studying its effects on text generation models performing various tasks, including translation, summarization, and language modelling. Their model hijacking attack, called Ditto, does not involve adding any triggers or modifying input, which means that the attack remains fully hidden after the target model is deployed. In other words, all inputs received by the model are benign inputs. The attack works by first gathering a set of tokens for each label in the hijacking dataset. The attacker then gives these tokens to the model and checks the results. Once the attacker knows how the model responds to these tokens, they can use a masked language model to manipulate the outputs. After the model is hijacked, Ditto checks the model's output against different token sets to determine the label. The researchers investigated using the ONION defense [108] to detect and remove hijacked data points (instead of outliers). However, it appears that this defense is not a foolproof solution against the Ditto attack. In general, most of the proposed defenses against this attack, such as input validation and filtering, need further investigation in order to reduce legal risks for model providers.

**Member Inference Attack (MIA).** ML models tend to memorize sensitive information, which puts them at risk of being targeted by MIA. These attacks involve an attacker attempting to determine whether an input sample was used to train the model [111, 112, 114]. Research has shown that larger language models, like GPT models, have a higher tendency to memorize training data when effectively prompted. As model capacity, frequency of examples in the training dataset, and prompt tokens increase, the risk of memorization also increases[113]. Table 7 shows some important research of MIA.

Table 7. Some studied member inference attacks

| Ref. | Attack Explanation |
|---|---|
| [114] | An attacker can use input-output pairs from the LLMs to train a binary classifier that can determine whether a specific individual was part of the model's training dataset. This is done by providing inputs representative of that individual and checking the output [10]. A widely used approach for training a binary classifier-based MIA is the work of Shokri et al [114]. |
| [115] | Hisamoto et al. investigated MIA for sequence-to-sequence models such as machine translation models (MT). The researchers created a dataset using MT and used Shokri et al.'s classifier to test this privacy attack. The study found that sentence-level membership in these models was hard to determine for attackers. Nevertheless, the models still have a risk of leaking private information. |
| [120] | Duan et al. demonstrated that a MIA can effectively infer the prompt used to generate responses from prompted models like GPT-2. They found that prompted models are over five times more susceptible to privacy leakage compared to fine-tuned models. |
| [121] | Mattern et al. introduced the neighborhood attack which involves providing a target sample and utilizing a pretrained masked language model to generate neighboring sentences that are highly similar through techniques like word replacement. By comparing the model scores of these neighbors to the target sample's score, we can determine whether the target sample is a member of the training set. |

De-duplicating the training dataset [116], using DP training [84], adding regularization [117] and using machine unlearning method [118, 119] (intentionally modifying the ML models to forget specific data points or features to protect sensitive data and preventing it from being used for decision making or prediction) are some proposed methods to defend against MIA that need further studies.

## 4 Conclusion & Future Work

The rapid advancement of LLMs has revolutionized language processing, but also opened new avenues for attacks by malicious actors. In this survey, we categorized attacks into two groups: attacks on LLMs applications and attacks on the models themselves and explored the significant attack types in each category. We provided comprehensive definitions of these attacks and explored the latest research on their implementation and countermeasures. We used our designed prompts to assess the impact and potential consequences of these attacks in some cases. Additionally, we explored real-life scenarios to better understand the practical implications and risks associated with these attacks. These attacks can have serious consequences, compromising the privacy and security of users' data, and undermining the reliability and trustworthiness of large language models. As future work, we plan to introduce a framework to evaluate the resilience of LLMs-integrated applications against prompt injection attacks. Additionally, we aim to investigate the feasibility of launching various attacks on the system message of LLM-integrated virtual assistants. By shedding light on these challenges, we hope to inspire researchers and developers to further explore and address these issues in their future works.

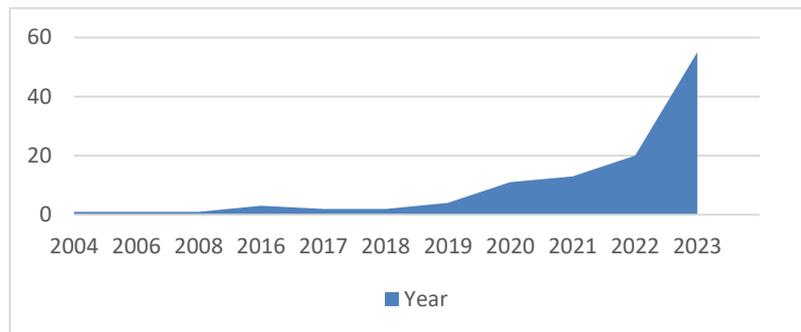

**Fig. 3.** Number of papers surveyed by year of publication. The rapid growth of works in the field of LLMs in 2023 underscores the increasing importance and prevalence of these models.

**Acknowledgments.** The authors would like to thank the Logistics and Supply Chain MultiTech R&D Centre, Hong Kong for providing the support to this work.